\documentclass[aps,prl,superscriptaddress,floatfix,twocolumn]{revtex4-1}
\usepackage{graphicx,graphics}
\usepackage{dcolumn}
\usepackage{amsmath,amssymb,amsfonts}
\usepackage{latexsym,verbatim}
\usepackage{bm}
\usepackage{color}
\usepackage{ulem}
\usepackage[breaklinks=true,colorlinks,citecolor=blue,linkcolor=blue,urlcolor=blue]{hyperref}

\def \Im {\mathop {\rm Im}}

\usepackage{bibunits}

\begin{document}

\title{Confining graphene plasmons to the ultimate limit}

\author{Alessandro Principi}
\affiliation{School of Physics and Astronomy, University of Manchester, Manchester, M13 9PL, UK}

\author{Erik van Loon}
\affiliation{Radboud University, Institute for Molecules and Materials, NL-6525 AJ Nijmegen, The Netherlands}

\author{Marco Polini}
\address{Istituto Italiano di Tecnologia, Graphene Labs, Via Morego 30, I-16163 Genova, Italy}

\author{Mikhail I. Katsnelson}
\affiliation{Radboud University, Institute for Molecules and Materials, NL-6525 AJ Nijmegen, The Netherlands}

\begin{abstract}
Graphene plasmons have recently attracted a great deal of attention because of their tunability, long lifetime, and high degree of field confinement 
in the vertical direction. Nearby metal gates have been shown to modify the graphene plasmon dispersion and further confine their electric field.
We study the plasmons of a graphene sheet deposited on a metal, in the regime in which metal bands do not hybridize with massless Dirac fermion bands. 
We derive exact results for the dispersion and lifetime of the plasmons of such hybrid system, taking into account metal nonlocalities.
The graphene plasmon dispersion is found to be acoustic and pushed down in energy towards the upper boundary of the intraband graphene particle-hole continuum, thereby strongly enhancing the vertical confinement of these excitations. Landau damping of such acoustic plasmons due to particle-hole excitations in the metal gate is found to be surprisingly weak, with quality factors exceeding $Q = 10^{2}$.
\end{abstract}
\maketitle


\begin{bibunit}

{\it Introduction.}---Recently, the fundamental properties of graphene plasmons and hybrid plasmon-phonon polaritons in graphene encapsulated in hexagonal boron nitride (hBN) have been studied in an extensive manner~\cite{grigorenko_naturephoton_2012,basov_science_2016,low_naturemater_2017}. In encapsulated graphene, plasmons have record-high lifetimes approaching $1~{\rm ps}$ both in the mid-infrared~\cite{Woessner_nature_mater_2015} and Terahertz (THz)~\cite{Alonso_nature_nano_2017,lundeberg_science_2017} spectral ranges, while at the same time displaying strong vertical confinement. 

The plasmon dispersion relation in graphene can be engineered not only by coupling them to standing Fabry-P\'{e}rot phonon polariton modes of hBN slabs~\cite{tomadin_prl_2015}, but also by placing metal gates nearby. These have been shown to play two vital roles. On the one hand, when shaped in the form of split gates, the associated $p$-$n$ junctions allow to detect plasmons electrically thanks to the photothermoelectric effect~\cite{lundeberg_naturemater_2017}. On the other hand, they screen the long-range tail of the electron-electron interaction potential, yielding acoustic plasmon modes~\cite{Stauber_njp_2012,Gu_apl_2013,Principi_ssc_2011} whose associated electric field is tightly confined to the small volume between the metal gate and graphene~\cite{Alonso_nature_nano_2017}. The latter is typically filled with an hBN spacer, which can be thinned down to a single layer or even removed. The modification of the plasmon dispersion from the usual unscreened $\sqrt{q}$ form~\cite{grigorenko_naturephoton_2012}, $q$ being the in-plane plasmon wave number, to the metal-screened form $\propto q$ at long wavelengths yields, for a fixed plasmon frequency $\omega$, modes with large $q$ and therefore strong vertical confinement $\propto q^{-1}$.

We also note in passing that hybrid graphene/metal structures may be of high technological relevance in the fields of gas and biological sensing~\cite{Salihoglu_apl_2012,Choi_opt_express_2011,rodrigo_science_2015}. Current devices for sensing usually contain periodic metallic structures, i.e.~gratings, deposited on graphene, which are used to couple far-field light to plasmons. Understanding the properties of graphene plasmons in the presence of a nearby metal is therefore 
of high technological relevance.

In this context, a natural question arises. What is the ultimate limit of vertical confinement for graphene plasmons? The answer seems to be that maximum vertical confinement can be achieved by depositing graphene {\it directly} on the metal gate. (Note that here we are not interested in the case of samples where graphene is grown by chemical vapour deposition on selected metals~\cite{Meng_jpcm_2012,Sun_prb_2011,Orofeo_carbon_2012,Generalov_carbon_2012,Cupolillo_surf_sci_2013,Kravets_sci_rep_2014,Politano_prb_2011,Politano_nanoscale_2014,Politano_2D_mater_2017}. Often, in this case, hybridization occurs between graphene and metal bands, leading to plasmonic excitations that share very little with graphene plasmons and that are usually accompanied by strong damping.) Superficially, however, this does not sound as a good choice. The point is that, naively speaking, plasmons in a graphene sheet deposited very close to a metal gate are expected to decay easily by emitting electron-hole pairs in the metal and therefore be strongly Landau damped. 

In this Article, we study the plasmons of a graphene sheet deposited on a metal, down to the ultimate limit of zero distance between the two. The metal is treated beyond the perfect-conductor approximation. At long wavelength, the problem can be solved exactly with the Wiener-Hopf method~\cite{Noble_WH_book} (see also Supplemental Information), in the spirit of Reuter and Sondheimer's work on the anomalous skin effect~\cite{Reuter_procroyal_1948}. Such a theory includes nonlocalities due to the finite electronic mean free path in the metal, and dielectric nonlocalities quantified by the metal screening function $\epsilon^{\rm M}(q,\omega)$. We show that graphene plasmons survive even in the zero-distance limit. The presence of metal gates proves to be an efficient way to manipulate the plasmon dispersion and reach ultra-high levels of field confinement. Moreover, by showing that the plasmon decay rate due to Landau damping enabled by the metal scales like $\sim q^2 \ln(q)$ at long wavelength and is numerically small, we also conclude that that these excitations remain well defined in the presence of metallic substrates. The theory we develop here is very general and can be used to describe the propagation of plasmons in the presence of different metal gates, provided no hybridization occurs between graphene and the metal.

{\it Model and general results.}---We consider a two-dimensional (2D) graphene sheet at a distance $d$ from the surface of a metal gate, which
is modeled as a three-dimensional electron gas (3DEG) occupying the half-space $z\leq 0$. The surface of the metal is assumed to be flat and the graphene sheet is placed at $z=d$.  The metal is characterized by the electronic density $n_{\rm M}$ and by a band-energy dispersion $\varepsilon_{{\bm k}, {\rm M}} = \hbar^2 {\bm k}^2/(2m_{\rm M})$, $m_{\rm M}$ being the effective mass of electrons in the metal. Therefore, the metal Fermi energy is $\varepsilon_{{\rm F}, {\rm M}} = \hbar^2 k_{{\rm F}, {\rm M}}^2/(2m_{\rm M})$, with $k_{{\rm F}, {\rm M}} = (3\pi^2 n)^{1/3}$~\cite{Giuliani_and_Vignale} the Fermi wave number. 
We also define the metal Fermi velocity, ${\bar v}_{\rm M} = \hbar k_{{\rm F}, {\rm M}}/m_{\rm M}$, the density-of-states at the Fermi energy, $\nu_{0, {\rm M}} = 3 n_{\rm M}/\varepsilon_{{\rm F}, {\rm M}}$, and the Thomas-Fermi screening wave number $q_{{\rm TF}, {\rm M}} = \sqrt{4\pi e^2 \nu_{0, {\rm M}}}$.
We treat electrons in graphene as massless Dirac fermions~\cite{Katsnelson_book}, characterized by a density-independent Fermi velocity $v_{\rm F}$. We assume the graphene sheet to be doped with an electron density $n$ above the Dirac point. In what follows $k_{\rm F}$ is the Fermi wave number in graphene and $\varepsilon_{\rm F}=\hbar v_{\rm F} k_{\rm F} >0$ its Fermi energy.

Neglecting retardation effects, the plasmon dispersion is found by solving Poisson's equation for the self-consistent electrostatic potential $\phi({\bm r},z,t)$,
${\bm \nabla}^2\phi({\bm r},z,t)= -4\pi \rho({\bm r},z,t)$, where the density $\rho({\bm r},z,t)$ is self-induced by $\phi({\bm r},z,t)$. ${\bm r}$ denotes the position of a point in a 2D plane parallel to the surface of the metal. Assuming translational and rotational symmetry in the 2D planes parallel to the metal surface, and taking the Fourier transform with respect to ${\bm r}$, Poisson's equation becomes
\begin{eqnarray} \label{eq:Maxwell_fund}
(\partial_z^2 - q^2) \phi_{q,\omega}(z) = -4\pi e \rho_{q,\omega}(z)
~,
\end{eqnarray}
where $\phi_{q,\omega}(z)$ and $\rho_{q,\omega}(z)$ are the Fourier transforms, respectively, of the self-consistent potential and number density, while $e$ is the electronic charge. The quantity $\rho_{q,\omega}(z)$ is rewritten as the sum of the individual contributions of the metal [$\rho_{{\rm M}, q,\omega}(z)$] and graphene sheet [$\rho_{{\rm G},q,\omega}$] as $\rho_{q,\omega}(z) = \rho_{M,q,\omega}(z) \theta(-z) + \rho_{G,q,\omega} \delta(z-d)$. In the linear-response regime, $\rho_{G,q,\omega} = \chi_{\rho\rho}(q,\omega) \phi_{q,\omega}(d)$, where $\chi_{\rho\rho}(q,\omega)$ is the noninteracting density-density linear response function of 2D massless Dirac fermions~\cite{Hwang_prb_2007}. 
We will use of the Boltzmann equation for the distribution function of electrons in the metal, $f_{q,\omega}({\bm k}, z)$, to calculate the density $\rho_{M,q,\omega}(z)$ induced by the field $\phi_{q,\omega}(z)$.

We solve the linear differential problem posed by Eq.~(\ref{eq:Maxwell_fund}) in the three separate regions $z>d$, $0<z<d$, and $z<0$. We impose the continuity of the potential at $z=0,d$, and of its derivative at $z=0$. At $z=d$ we also have the boundary condition (i.e.~discontinuity of the electric field perpendicular to graphene)
\begin{eqnarray}
\partial_z \phi_{q,\omega}(z) \big|_{d^-}^{d^+} = - 4 \pi e^2 \chi_{\rho\rho}(q,\omega) \phi_{q,\omega}(d)
~.
\end{eqnarray}
Solving the resulting linear differential problem, we find that for $\phi_{q,\omega}(z)$ to be non-zero we must 
satisfy
the {\it plasmon equation}
$\epsilon(q,\omega) \equiv 1 - V(q,\omega)\chi_{\rho\rho}(q,\omega) = 0$,
whose solution $\omega=\omega_{\rm p}(q)$ gives the plasmon dispersion in graphene in the presence of the metal gate.
Here $V(q,\omega) = 2 \pi e^2/\big[ q \epsilon^{\rm M}(q,\omega) \big]$ is the effective electron-electron interaction in graphene, and
\begin{eqnarray} \label{eq:effective_Coulomb_interaction}
\epsilon^{\rm M}(q,\omega) = \left( 1 + \frac{Z_{q,\omega} - 1}{Z_{q,\omega} + 1} e^{-2 q d} \right)^{-1}
\end{eqnarray}
is the metal screening function.
Here $Z_{q,\omega} \equiv q\phi_{q,\omega}(z)/\partial_z \phi_{q,\omega}(z)\big|_{z\to 0^-}$ plays the role of a ``dimensionless surface impedance'' and 
depends {\it only} on the properties of the metal.
Note that, in the limit $Z_{q,\omega}\to 1$, $V(q,\omega)$ reduces to the Coulomb interaction of an isolated graphene sheet~\cite{Hwang_prb_2007}. Conversely, in the limit $Z_{q,\omega}\to 0$ we recover the result for a graphene sheet in the presence of a perfect conductor~\cite{Alonso_nature_nano_2017,Principi_ssc_2011}. In what follows we briefly summarize 
how to calculate this quantity. 

To calculate $Z_{q,\omega}$ we need to determine the ratio $\phi_{q,\omega}(z)/\partial_z \phi_{q,\omega}(z)$ at $z\to 0^-$ or, equivalently, $\partial_z \phi_{q,\omega}(z)\big|_{z\to 0^-}$ as a function of $\phi_{q,\omega}(0)$. The latter quantity, momentarily unspecified, represents an alternative boundary condition that we can use to solve the Poisson's equation~(\ref{eq:Maxwell_fund}) for $z<0$. The quantity $\phi_{q,\omega}(0)$ accounts in fact for the distribution of charges in the half-space $z>0$ (included the graphene sheet). 
This fact allows us to employ a mathematical trick that dramatically simplifies our calculation. In fact, we can now introduce the new function ${\bar \phi}_{q,\omega}(z)$, which coincides with $\phi_{q,\omega}(z)$ for $z<0$ and equals $\phi_{q,\omega}(-z)$ for $z>0$. 
${\bar \phi}_{q,\omega}(z)$ is clearly an even function of $z$. We can then solve the Eq.~(\ref{eq:Maxwell_fund}) for ${\bar \phi}_{q,\omega}(z)$ with the boundary condition ${\bar \phi}_{q,\omega}(0)=\phi_{q,\omega}(0)$. By the uniqueness of the solution of the Poisson's equation, we are guaranteed that ${\bar \phi}_{q,\omega}(z)$ coincides with $\phi_{q,\omega}(z)$ for $z<0$, and therefore $Z_{q,\omega} = q{\bar \phi}_{q,\omega}(z)/\partial_z {\bar \phi}_{q,\omega}(z)\big|_{z\to 0^-}$.

For the sake of brevity, from now on we omit the subscript ${\rm M}$ when referring to the properties of the metal. This should not generate any confusion, since all quantities in the following equations [(\ref{eq:Boltzmann_RTA})-(\ref{eq:kappa_defs})] refer to the metal. Only when strictly necessary, we will reinstate the subscript ${\rm M}$. 
We consider the following Boltzmann equation for the distribution function $f_{\bm k}\equiv f_{\bm k}({\bm r},z, t)$:
\begin{eqnarray} \label{eq:Boltzmann_RTA}
\partial_t f_{\bm k} + e {\bm E} \cdot {\bm \nabla}_{\bm k} f_{\bm k} + {\bm v}_{\bm k}\cdot {\bm \nabla}_{\bm r} f_{\bm k} = - \tau^{-1}(f_{\bm k} - f^{0}_{{\bm k}})
~,
\end{eqnarray}
where ${\bm E} \equiv -{\bm \nabla}\phi({\bm r},z,t)$ is the self-induced electric field, ${\bm v}_{\bm k} = \hbar {\bm k}/m$ is the 3D particle velocity and $f^{0}_{{\bm k}} = [e^{\beta(\varepsilon_{\bm k}-\mu)} + 1]^{-1}$ is the equilibrium Fermi-Dirac distribution. Finally, $\tau$ 
models
a finite transport lifetime in the metal. Setting $f_{\bm k}({\bm r},z,t) = f^{0}_{{\bm k}} + \delta f_{{\bm k},q,\omega}(z) e^{i(qx+\omega t)}$ and $\phi({\bm r},z,t) = \phi_{q,\omega}(z) e^{i (qx+\omega t)}$, and linearizing Eq.~(\ref{eq:Boltzmann_RTA}) with respect to $\delta f_{{\bm k},q,\omega}(z)$ and $\phi_{q,\omega}(z)$ we get
(in the remainder of the section we 
omit
the variables $q$ and $\omega$)
\begin{eqnarray} \label{eq:Boltzmann_RTA_2}
v_{{\bm k}}^z \big[ A_{\bm k}  \delta f_{\bm k}(z) + \partial_z \delta f_{\bm k}(z) \big] = e \partial_\varepsilon f^{0}_{{\bm k}} \Phi_{\bm k}(z)
~,
\end{eqnarray}
where $\varepsilon = \varepsilon_{\bm k}$, $\Phi_{\bm k}(z) \equiv i q v_{{\bm k}}^x \phi(z)+v_{{\bm k}}^z \partial_z \phi(z)$, 
$A_{\bm k}  = (i\omega +\tau^{-1} + i q v_{{\bm k}}^{x})/v_{{\bm k}}^{z}$, and $v_{{\bm k}}^{\alpha}$ with $\alpha=x,z$ being the components of the velocity ${\bm v}_{\bm k}$.
The general solution of this equation is (recall that $z<0$)
\begin{equation} \label{eq:deltaf_general}
\delta f_{\bm k}(z) 
=
e^{- A_{\bm k}  z} \Bigg[ F({\bm v}_{\bm k}) 
+ \frac{e\partial_\varepsilon f^{0}_{{\bm k}}}{v_{\bm k}^z} \int_{-\infty}^z d\zeta e^{A_{\bm k} \zeta}
\Phi_{\bm k}(\zeta)
\Bigg]
~.
\end{equation}
We now determine $F({\bm v}_{\bm k})$ for the two cases, $v_{{\bm k}}^{z}>0$ and $v_{{\bm k}}^{z}<0$, separately. 
When $v_{{\bm k}}^{z}>0$, $\delta f_{{\bm k}}(z)$ describes particles traveling from deep inside the metal towards its surface. In this case, we choose $F({\bm v}_{\bm k})$ in such a way that $\delta f_{\bm k}(z)$ (and hence metallic properties deep inside the bulk) does not diverge for $z\to -\infty$.
This implies that $F({\bm v}_{\bm k}) = 0$ for $v_{{\bm k}}^{z}>0$, and
\begin{equation} \label{eq:deltaf_plus}
\delta f_{\bm k}^+(z) = \frac{e\partial_\varepsilon f^{0}_{\bm k}}{v_{\bm k}^z} \int_{-\infty}^{z} d\zeta 
\Phi_{\bm k}(\zeta)
e^{A_{\bm k} (\zeta - z)}
~,
\end{equation}
where the superscript ``$+$'' recalls the fact that here $v_{{\bm k}}^{z}>0$.
Here we replaced $\phi(\zeta) \to {\bar \phi}(-\zeta)$, since $z<0$.
When $v_{{\bm k}}^{z}<0$ we consider the scattering at the boundary to be specular with probability $p$ and diffusive with probability $1-p$. The assumption $p={\rm const}$ is required to solve the problem analytically~\cite{Reuter_procroyal_1948}. In realistic models of surface scattering $p$ is strongly dependent on the incidence angle
~\cite{Okulov_sjltp_1979,Falkovsky_adv_phys_1983}. Under this assumption,
$\delta f^-_{\bm k}(z=0) = p \delta f^+_{\bm k}(z=0)\big|_{v_z\to -v_z}$,
where $\delta f^-_{\bm k}(z)$ describes electrons moving with velocity $v_{{\bm k}}^{z}<0$. This, in turn, implies
\begin{eqnarray} \label{eq:deltaf_minus}
\delta f_{\bm k}^-(z) &=& 
\frac{e \partial_\varepsilon f^0_{\bm k}}{|v_{\bm k}^z|} 
\int_{z}^\infty d\zeta \Theta^{(p)}_\zeta 
\Phi_{\bm k}(\zeta)
e^{A_{\bm k} (\zeta-z)}
~,
\end{eqnarray}
where $\Theta^{(p)}_\zeta = p + (1-p) \Theta(-\zeta)$, $\Theta(\zeta)$ being the usual Heaviside step function.
From $\delta f_{\bm k}(z) = \Theta(v_{\bm k}^z) f^+_{\bm k}(z) + \Theta(-v_{\bm k}^z) f^-_{\bm k}(z)$ we can calculate the density of the metal as $\rho(z) = \sum_{\bm k} \delta f_{\bm k}(z)$. After some algebra, we find
\begin{equation} \label{eq:J_metal_final_2}
\rho(z) =
e \nu_{0}
\int_{-\infty}^\infty d\zeta \big[\Theta^{(p)}_\zeta \kappa(z-\zeta) {\bar \phi}(\zeta) 
+
(1-p) \kappa^z(z) {\bar \phi}(0)
\big]
~,
\end{equation}
where $\kappa(z) = {\tilde \kappa}(z) - \delta(z)$, and
\begin{equation} \label{eq:kappa_defs}
{\tilde \kappa}(z) = \frac{i\omega+1/\tau}{2 {\bar v}} \int_0^{\pi/2} d\theta \frac{\sin(\theta)}{\cos(\theta)} e^{ - B_\theta |z| }
J_0\big[q |z| \tan(\theta)\big]
~,
\end{equation}
Here $B_\theta \equiv (i\omega +1/\tau)/[{\bar v} \cos(\theta)]$ and $\theta$ is the angle between ${\bm k}$ and the ${\hat {\bm z}}$ axis. $\kappa^z(z)$ is not relevant here, and is given in the Supplemental Material~\cite{SOM}, where we provide analytical results for the case $p=0$. Here we
analyze the case $p=1$ (perfect reflection at the interface) providing both analytical and numerical results.

{\it Analytical results for $p=1$}---In this case, Poisson's equation is rewritten as
\begin{equation} \label{eq:p1_system}
(\partial_z^2 - q^2) {\bar \phi}(z) = -q_{{\rm TF},M}^2 \int_{-\infty}^\infty d\zeta \kappa_{q,\omega}(z-\zeta) {\bar \phi}(\zeta)~,
\end{equation}
where ${\bar \phi}(z)$ is continuous everywhere, while $\partial_z {\bar \phi}(z)$ has a jump for $z=0$. 
Solving Eq.~(\ref{eq:p1_system}) we find
\begin{equation} \label{eq:Z_p_1}
Z_{q,\omega} = \int_{0}^{+\infty} \frac{dq_z}{\pi} \frac{2 q}{q_z^2+q^2-q_{{\rm TF},M}^2\kappa_{q,\omega}(q_z)}
~.
\end{equation}
Setting $\omega = c_{\rm p} q$ in 
the plasmon equation $\epsilon(q,\omega) = 0$
and taking the $q\to 0$ limit (neglecting the imaginary part) we find $Z_{q,\omega} = q/q_{{\rm TF}, {\rm M}}$ and the acoustic-plasmon dispersion
\begin{eqnarray} \label{eq:plasmon_velocity}
\omega_{\rm p}(q) = v_{\rm F} q \frac{1+\Xi}{\sqrt{1+2 \Xi}}~,
\end{eqnarray}
where $\Xi = 4\pi e^2 (d+q_{{\rm TF},M}^{-1}) \nu(\varepsilon_{\rm F})$ and $\nu(\varepsilon_{\rm F}) = N_{\rm F} \varepsilon_{\rm F}/(2\pi \hbar^2 v_{\rm F}^2)$ is the density of states of graphene at the Fermi energy. 
Eq.~(\ref{eq:plasmon_velocity}) allows to define the acoustic-plasmon group velocity $c_{\rm p} = v_{\rm F} (1+\Xi)/\sqrt{1+2 \Xi}>v_{\rm F}$.

\begin{figure}[t]
\centering
\includegraphics{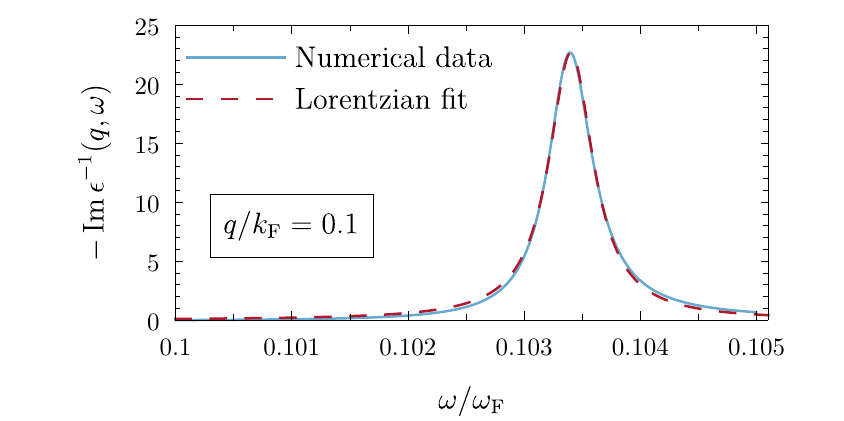}
\caption{The loss function of a graphene sheet on a metal gate ($p=1$). Solid line: the quantity $-{\rm Im}[\epsilon^{-1}(q,\omega)]$ (for a color plot see Fig.~\ref{fig:two}) is plotted as a function of $\omega$ (in units of $\omega_{\rm F} = \varepsilon_{\rm F}/\hbar$) for a fixed value of $q=0.1~k_{\rm F}$. Dashed line: a Lorentzian fit used to determine the lifetime. 
\label{fig:one}}
\end{figure}

Let us now derive the plasmon lifetime $\tau_{\rm p}(q)$~\cite{SOM}. We set $\omega = \omega_{\rm p}(q) + i/[2 \tau_{\rm p}(q)]$ in 
the plasmon equation.
Expanding it for small $\tau_{\rm p}^{-1}(q)$ 
and for $q\to 0$,
we get
\begin{align}\label{eq:inv_lifetime}
\frac{1}{\tau_{\rm p}} \to \frac{c_{\rm p}  (c^2_{\rm p} - v^2_{\rm F})}{v_{\rm F}^2{\bar v}_{\rm M} q_{{\rm TF}, {\rm M}}} \frac{c_{\rm p} - \sqrt{c^2_{\rm p} - v_{\rm F}^2}}{1+d q_{{\rm TF}, {\rm M}}}
q^2  \left[2\ln\left(q_{{\rm TF}, {\rm M}}/q\right)-K\right]
\end{align}
where
\begin{align}
K=\ln\left(c^2_{\rm p}/{\bar v}^2_{\rm M}-1\right) + \frac{(\pi^2/4-1)c^2_{\rm p} -{\bar v}^2_{\rm M}/2}{c^2_{\rm p} - {\bar v}^2_{\rm M}}~. 
\end{align}
This dependence of the damping on wave vector $q$, parametrically smaller than the plasmon frequency, shows that acoustic plasmons are long-lived excitations.

{\it Numerical results and discussion.}---We numerically calculate and fit the loss function~\cite{Giuliani_and_Vignale,eels} $-\Im [\epsilon^{-1}(q,\omega)]$ for the case $d=0$. 
We compare these results with the analytical ones of Eqs.~(\ref{eq:plasmon_velocity}) and~(\ref{eq:inv_lifetime}). We find that acoustic plasmons in graphene are pushed towards the boundary of the particle-hole continuum $\omega = v_{\rm F} q$ by the screening exerted by the metal. This fact implies that they become extremely localized in the direction perpendicular to graphene, at much lower frequencies than those of graphene/dielectric stacks. 

\begin{figure}[t]
\centering
\includegraphics[width=0.99\columnwidth]{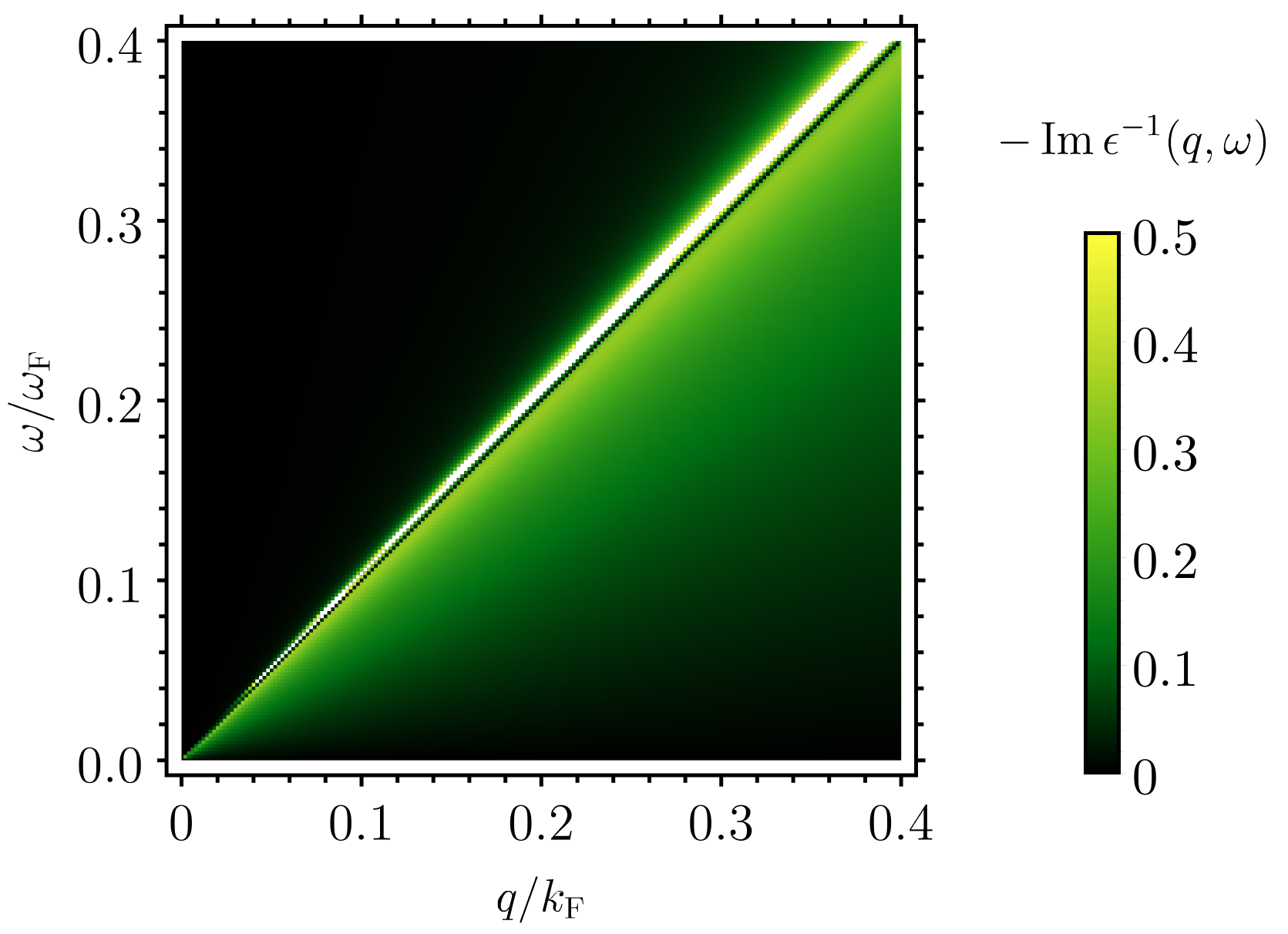}
\caption{
A color plot of the loss function $-\Im[\epsilon^{-1}(q,\omega)]$ as a function of $q/k_{\rm F}$ and $\omega/\omega_{\rm F}$. A sharp acoustic plasmon mode is visible just above the upper edge of the graphene intraband particle-hole continuum, $\omega = v_{\rm F} q$. 
\label{fig:two}}
\end{figure}

We have performed numerical calculations for the following electron densities, $n_{\rm M}= 10^{21}~{\rm cm}^{-3}$ and $n=10^{12}~{\rm cm}^{-2}$. For these parameters, the acoustic plasmon group velocity is 
$c_{\rm p}\approx 1.04~v_{\rm F}$. The Fermi velocity of the metal is 35\% of $v_{\rm F}$. 
Fig.~\ref{fig:one} shows the loss function~\cite{Giuliani_and_Vignale,eels}
plotted as a function of $\omega$ (in units of $\omega_{\rm F}=\varepsilon_{\rm F}/\hbar$) and for a fixed value of $q=0.1~k_{\rm F}$. 
A plasmon mode is clearly visible, in the form of a Lorentzian peak, centered at a frequency slightly above the particle-hole-continuum threshold $\omega=v_{\rm F} q$. In Fig.~\ref{fig:two} we show a 2D color plot of the same quantity, as a function of $q/k_{\rm F}$ and $\omega/\omega_{\rm F}$. A linearly-dispersive plasmon mode can be easily recognized at energies slightly above the upper edge of the interband particle-hole continuum ($\omega = v_{\rm F} q$). In spite of the damping introduced by the metal, the plasmon dispersion remains extremely sharp. 
Moreover,
the acoustic plasmon carries a much larger spectral weight of that carried by the intraband particle-hole excitations. We therefore conclude that surface-science techniques such as electron-energy loss spectroscopy~\cite{eels} or scattering-type near-field optical spectroscopy~\cite{grigorenko_naturephoton_2012,basov_science_2016,low_naturemater_2017} can efficiently probe this acoustic plasmon mode and distinguish it from the incoherent continuum of particle-hole excitations.

\begin{figure}[t]
\centering
\includegraphics{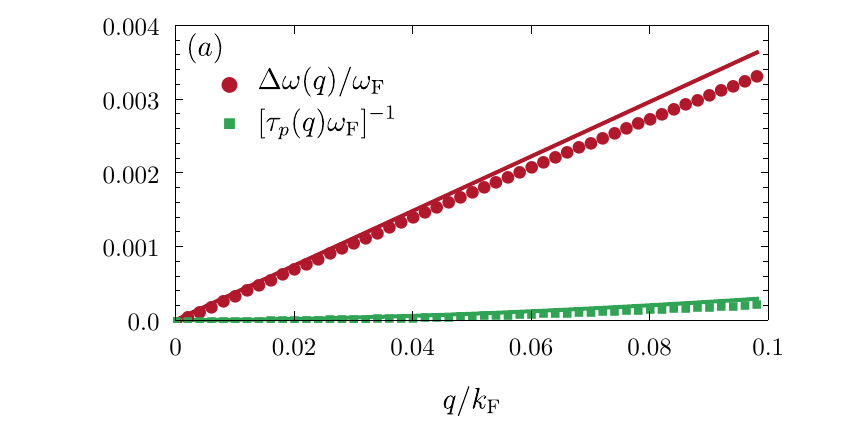}
\includegraphics{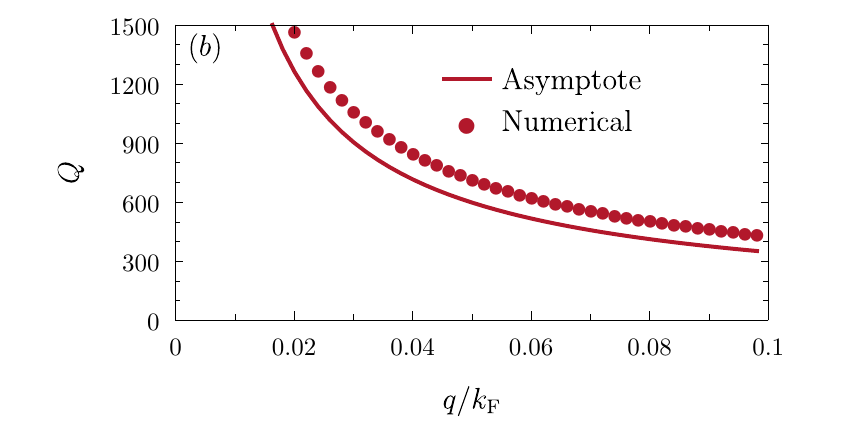}
\caption{
Plasmon dispersion, lifetime and quality factor. (a) Plasmon dispersion relation and lifetime. The red circles show the normalized distance $\Delta\omega_{\rm p}(q)/\omega_{\rm F}$ of the acoustic plasmon dispersion from the upper bound of the intraband particle-hole continuum,
as obtained from Lorentzian fits to the numerical data. 
The solid red line shows the same quantity as from the analytical result in Eq.~(\ref{eq:plasmon_velocity}). The green squares show the normalized width $[\tau_{\rm p}(q)\omega_{\rm F}]^{-1}$ of the plasmon obtained from the half-width at half-maximum of the Lorentzians, while the green line shows the analytical result in Eq.~(\ref{eq:inv_lifetime}). 
(b) The acoustic plasmon quality factor $Q = \omega_{\rm p}(q) \tau_{\rm p}(q)$ is plotted as a function of $q/k_{\rm F}$.
\label{fig:three}}
\end{figure}

In Fig.~\ref{fig:three}(a) we compare the numerical results of 
the
fitting procedure with the analytical asymptotic expressions given in Eqs.~(\ref{eq:plasmon_velocity}) and~(\ref{eq:inv_lifetime}). At small $q$, a good match is found. The width $\tau^{-1}_{\rm p}(q)$ of the plasmon peak is always much smaller than the distance in energy between the plasmon and the upper boundary of the particle-hole continuum [$\Delta \omega(q) = \omega_{\rm p}(q) -  v_{\rm F}q$], showing that the plasmon remains well-defined and extremely sharp.
Note also that subleading corrections to the formula~(\ref{eq:plasmon_velocity}) introduce only small deviations of the plasmon dispersion from linearity. In fact, comparing $\Delta \omega(q)$ given in Fig.~\ref{fig:three}(a) with the plasmon frequency $\omega_{\rm p}(q)$ extracted from Fig.~\ref{fig:two} we see that $\Delta \omega(q)/\omega_{\rm p}(q)\lesssim 3\%$. 
Therefore, the plasmon remains acoustic to a very good approximation in a wide range of momenta. 
Finally, in Fig.~\ref{fig:three}(b) we show the quality factor $Q = \omega_{\rm p}(q) \tau_{\rm p}(q)$ as a function of the wavevector $q$ (in units of the Fermi wavevector $k_{\rm F}$). The strong suppression of the metal-induced lifetime at small momenta leads to astonishingly large values of $Q$. This allows us to conclude that plasmon lifetimes will rather be limited by the same extrinsic effects (like impurities or phonons) present in graphene/dielectrics heterostructures. 
We can expect graphene on metal devices to exhibit figures of merit similar to other devices, with the added value of an ultra-strong vertical confinement, even in the THz range. 

{\it Conclusions.}---In this work we theoretically studied the plasmons of graphene 
on a metal substrate. We calculated their dispersion and intrinsic lifetime, showing (i) that they acquire an acoustic dispersion because of the screening exerted by the metal, and (ii) that their vertical confinement is greatly enhanced, when compared with that in samples on dielectric substrates. Finally, (iii) we proved that acoustic plasmons remain well defined excitations, even in the presence of the metal, since their damping rate is always parametrically and numerically much smaller than their energy. 
Although acoustic plasmons in graphene are pushed by the presence of the metal towards the upper bound of the intraband particle-hole continuum, their width remains so sharp that they are well separated from it in a wide range of wavevectors.

Even more interestingly, our work allows to extract plasmon lifetimes and figures of merit due to  Landau damping in the metal that are much larger than what observed experimentally~\cite{Politano_prb_2011}. In fact, we predict that, in the absence of extrinsic effects like grain boundaries, disorder or phonons, quality factors larger than $\sim 600$ can be achieved.
This allows us to conclude that current experiments are far away from the intrinsic regime, and that the short plasmon lifetimes that are observed should be attributed to extrinsic mechanisms.

{\it Acknowledgements.}---A.P., E.v.L., and M.I.K. acknowledge support from the ERC Advanced Grant 338957 FEMTO/NANO and from the NWO via the Spinoza Prize. M.P. is supported by the European Union's Horizon 2020 research and innovation programme under grant agreement No.~696656 ``GrapheneCore1''.

\end{bibunit}

\clearpage

\onecolumngrid

\appendix

\section{Fourier transforms of $\kappa(z)$ and $\kappa_z(z)$}
\label{app:FT_kappa}
We consider
\begin{eqnarray}
&& {\kappa}_{q,\omega}(z) = \frac{i\omega+1/\tau}{2{\bar v}_M}  \int_0^{\pi/2} d\theta \frac{\sin(\theta)}{\cos(\theta)} \exp\left[ -\frac{i\omega +1/\tau}{{\bar v}_M \cos(\theta)} |z| \right] J_0\big[q |z| \tan(\theta)\big] - \delta(z)
~,
\end{eqnarray}
and we calculate its Fourier transform. 
Defining ${\tilde \omega} = \omega - i/\tau$, we get
\begin{eqnarray}
{ \kappa}_{q,\omega}(q_z) &=&
- \left\{ \frac{{\tilde \omega}}{2{\bar v}_M\sqrt{q_{z}^2+q^2}} \ln\left[ \frac{{\tilde \omega} -{\bar v}_M\sqrt{q_z^2+q^2}}{{\tilde \omega} + {\bar v}_M\sqrt{q_z^2+q^2}} \right] + 1 \right\}
~.
\end{eqnarray}
Note that $\kappa_{q,\omega}(q_z) \to 0$ as $q,q_z \to 0$ (as required by the gauge invariance). Moreover, expanding the function $\nu_{0,M} \kappa_{q,\omega}(q_z)/2$ for small $q,q_z$ we get
\begin{eqnarray}
\nu_{0,M} \kappa_{q,\omega}(q_z) \to \frac{n_M}{m_M} \frac{q^2+q_z^2}{\omega^2}
~,
\end{eqnarray}
as expected, since this function is nothing but the small-$q$ expansion of the density-density response function of the metal. As the same time, for ${\tilde \omega} \to 0$
\begin{eqnarray}
\nu_{0,M} \kappa_{q,\omega}(q_z) \to  -1 - i\frac{\pi{\tilde \omega}}{2{\bar v}_M\sqrt{q_{z}^2+q^2}}
~.
\end{eqnarray}
The small-frequency and -momentum behavior of the density-density response function $\nu_{0,M} \kappa_{q,\omega}(q_z)$ agrees with that of that of a 3DEG.~\cite{Giuliani_and_Vignale} Finally, in the limit $\tau\to\infty$,
\begin{eqnarray}
\kappa_{q,\omega}(q_z) =
- \Bigg\{ \frac{\omega}{2{\bar v}_M\sqrt{q_{z}^2+q^2}} \ln\Bigg| \frac{\omega -{\bar v}_M\sqrt{q_z^2+q^2}}{\omega + {\bar v}_M\sqrt{q_z^2+q^2}} \Bigg| + 1 \Bigg\} - i \frac{\pi \omega}{2{\bar v}_M\sqrt{q_{z}^2+q^2}} \Theta\big[{\bar v}_M^2 (q_z^2+q^2)- \omega^2\big]
~.
\end{eqnarray}

Similarly, we consider
\begin{eqnarray}
&& {\kappa}_{q,\omega}^z(z) =  \frac{1}{2} \int_0^{\pi/2} d\theta \sin(\theta) \exp\left[ -\frac{i\omega +1/\tau}{{\bar v}_M \cos(\theta)} |z| \right] J_0\big[q z \tan(\theta)\big] {\rm sign}(z)
~,
\end{eqnarray}
whose Fourier transform reads
\begin{eqnarray}
{ \kappa}^z_{q,\omega}(q_z) &=&
-\frac{i q_z}{q_z^2 + q^2} \kappa(q_z)
~.
\end{eqnarray}

\section{Case $p=1$: the real and imaginary parts of the potential for small $q$}
We consider Eq.~(\ref{eq:Z_p_1}), which we rewrite as
\begin{eqnarray} \label{eq:plasmon_final_app}
Z_{q,\omega} = \frac{2 q}{\pi} \int_{0}^{+\infty} \frac{dq_z}{q_z^2+q^2-q_{{\rm TF},M}^2\kappa_{q,\omega}(q_z)}
~,
\end{eqnarray}
using the symmetry properties of $\kappa_{q,\omega}(q_z)$. Setting $\omega = c_{\rm p} q$, in the limit $q\to 0$ we get
\begin{eqnarray}
\kappa_{q,\omega}(q_z) \to -1 - i \frac{\pi \omega}{2 {\bar v}_M q_z} \Theta({\bar v}_M^2 q_z^2 + {\bar v}_M^2 q^2 - \omega^2)
~,
\end{eqnarray}
which allows us to calculate
\begin{eqnarray} \label{eq:plasmon_final_app_Re}
Z_{q,\omega} &\to& \frac{2q}{\pi} \int_{0}^{+\infty}  \frac{dq_z}{q_z^2 + q_{{\rm TF},M}^2} \left[ 1 - i \frac{\pi q_{{\rm TF},M}^2 \omega}{2 {\bar v}_M q_z} \frac{\Theta({\bar v}_M^2 q_z^2 + {\bar v}_M^2 q^2 - \omega^2)}{q_z^2 + q_{{\rm TF},M}^2} \right]
\nonumber\\
&\to&
\frac{q}{q_{{\rm TF},M}} \left[ 1 + i \frac{\omega}{2 v q_{{\rm TF},M}} \ln \left(\frac{\omega^2 - {\bar v}_M^2 q^2}{{\bar v}_M^2 q_{{\rm TF},M}^2}\right) \right]
~.
\end{eqnarray}

\section{Case $p=0$}
In this case the Laplace equation for the electrostatic potential becomes
\begin{eqnarray} \label{eq:app_poisson_p0}
(\partial_z^2 - q^2) {\bar \phi}_{q,\omega}(z) &=& -q_{{\rm TF}, M}^2 \left[  \int_{0}^\infty d\zeta \kappa_{q,\omega}(z-\zeta) {\bar \phi}_{q,\omega}(\zeta) + \kappa^z_{q,\omega}(z) {\bar \phi}_{q,\omega}(0^-) \right]
~,
\end{eqnarray}
for $z<0$.
This equation can be solved by using the Wiener-Hopf method as we detail in what follows, similarly to the problem of the anomalous skin effect treated in Ref.~\onlinecite{Reuter_procroyal_1948}.

Let us first define the functions
\begin{eqnarray}
&& f(z) = \Theta(-z) \phi_{q,\omega}(z)
~,
\nonumber\\
&& g(z) = q_{{\rm TF},M}^2 \Theta(z) \left[  \int_{0}^\infty d\zeta \kappa_{q,\omega}(z-\zeta) {\bar \phi}_{q,\omega}(\zeta) + \kappa^z_{q,\omega}(z) {\bar \phi}_{q,\omega}(0^-) \right]
~,
\end{eqnarray}
which allow to rewrite Eq.~(\ref{eq:app_poisson_p0}) as
\begin{eqnarray} \label{eq:WH_1}
g(z) = (\partial_z^2 - q^2) f(z) + q_{{\rm TF},M}^2  \left[ \int_{-\infty }^\infty d\zeta \kappa_{q,\omega}(\zeta-z) f(\zeta) + \kappa^z_{q,\omega}(z) f(0) \right]
~.
\end{eqnarray}
Hereafter $f(0) \equiv f(0^-)$. Similarly for $f'(0) \equiv \partial_z f(z)\big|_{z\to 0^-}$. 
The form of Eq.~(\ref{eq:WH_1}) is suitable to apply the Wiener-Hopf technique. We will momentarily take its Fourier transform. For a generic function $A(z)$, we define it as
\begin{eqnarray}
A(q_z) &\equiv& {\cal F}[A(z)](q_z)
\nonumber\\
&=&
\int_{-\infty}^{\infty} dz e^{iq_z z} A(z)
~.
\end{eqnarray}
We will extend $q_z$ to the whole complex plane, and make use of theorems of complex analysis. For the reader convenience, we recall them here:~\cite{Noble_WH_book}
\begin{enumerate}

\item If $A(z)$ is such that $|A(z)|< c_1 e^{a z}$ for $z\to +\infty$, and $|A(z)| < c_2 e^{b z}$ for $z\to -\infty$, with $b>a$, then its Fourier transform $A(q_z)$ is analytic in the strip $a  < \Im(q_z) < b$.

\item Given $a$ and $b$, with $a<b$, if two functions $A(q_z)$ and $B(q_z)$ are analytical, respectively, for $\Im(q_z) > a$ and $\Im(q_z)<b$, and satisfy $A(q_z) = B(q_z)$ for $a< \Im(q_z) < b$, then there exist a unique function $C(q_z)$ analytical everywhere which coincides with $A(q_z)$ [$B(q_z)$] for $\Im(q_z) > a$ [$\Im(q_z)<b$].

\item Given ${\tilde A}$ and $p\geq 0$ constants (with $p$ an integer), if $A(q_z)$ is an integral function such that $|A(q_z)| \leq {\tilde A} |q_z|^p$ for $|q_z| \to \infty$, then $A(q_z)$ is a polynomial of degree $\leq p$. 

\item Assume $A(q_z)$ to be an analytic function in the strip $a<\Im(q_z)<b$ such that, for $q_z$ in the strip and $|\Re e(q_z)|\to \infty$, $|A(q_z)| < {\tilde A} |q_z|^{-p}$ (with $p>0$ and ${\tilde A}$ a constant). Then $A(q_z)$ can be written as
\begin{eqnarray} \label{eq:splitting_sum}
A(q_z) = A_+(q_z) + A_-(q_z)
~,
\end{eqnarray}
where $A_+(q_z)$ is regular for $\Im(q_z)>a$ and $A_-(q_z)$ is regular for $\Im(q_z) < b$, and
\begin{eqnarray}
A_\pm(q_z) = \pm \int_{i c_\pm -\infty}^{i c_\pm + \infty} \frac{d{\tilde q}_z}{2\pi i} \frac{A(q_z)}{{\tilde q}_z-q_z}
~,
\end{eqnarray}
for any $c_+$ and $c_-$ such that $a<c_+<\Im(q_z)<c_-<b$.

\item Assume $B(q_z)$ to be analytic and {\it different from zero} in the strip $a<\Im(q_z)<b$ and such that, for $q_z$ in the strip and $|\Re e(q_z)|\to \infty$, $|B(q_z)| \to 1$. Then $B(q_z)$ can be written as
\begin{eqnarray} \label{eq:splitting_product}
B(q_z) = \frac{B_+(q_z)}{B_-(q_z)}
~,
\end{eqnarray}
where $B_+(q_z)$ is regular for $\Im(q_z)>a$ and $B_-(q_z)$ is regular for $\Im(q_z) < b$, and
\begin{eqnarray} \label{eq:def_B_plus_minus}
B_\pm(q_z) = \exp\left[  \int_{i c_\pm -\infty}^{i c_\pm + \infty} \frac{d{\tilde q}_z}{2\pi i} \frac{\ln B(q_z)}{{\tilde q}_z-q_z} \right]
~,
\end{eqnarray}
for any $c_+$ and $c_-$ such that $a<c_+<\Im(q_z)<c_-<b$. This theorem is a corollary of the previous one, when the latter is applied to the function $A(q_z) = \ln B(q_z)$.

\end{enumerate}

Accordingly, we now observe that the Fourier transforms of the functions $\kappa_{q,\omega}(z)$ and $\kappa_{q,\omega}^z(z)$ [{\it i.e.} $\kappa_{q,\omega}(q_z)$ and $\kappa_{q,\omega}^z(q_z)$] have the same analytic properties, {\it i.e} they are analytic in the strip $|\Im q_z| < {\bar q}_z$, where ${\bar q}_z = \min\big[q,|\Im \sqrt{(\omega+i/\tau)^2-q^2}|\big]$. By definition, $g(q_z)$ is regular for $\Im q_z > -{\bar q}_z$, while $f(q_z)$ is analytic for $\Im q_z < {\bar q}_z$.
Let us consider the Fourier transform of $\partial_z^2 f(z)$ which, taking into account the discontinuities in $z=0$, reads
\begin{eqnarray}
{\cal F}[\partial_z^2 f(z)] = -q_z^2 f(q_z) + i q_z f(0) - f'(0)
~.
\end{eqnarray}
The function ${\cal F}[\partial_z^2 f(z)]$ must be bounded for $|q_z|\to \infty$. Therefore,
\begin{eqnarray} \label{eq:asymptotics_f}
f(q_z) \to i\frac{f(0)}{q_{z}} - \frac{f'(0)}{q_z^2}
~,
\end{eqnarray}
for $|q_z|\to \infty$. 

We now take the Fourier transform of Eq.~(\ref{eq:WH_1}) in the strip $|\Im(q_z)| < {\bar q}_z$. It reads
\begin{eqnarray} \label{eq:WH_2}
g(q_z) = -(q_z^2 + q^2) f(q_z) + i q_z f(0) - f'(0) + q_{{\rm TF},M}^2 \left[ \kappa_{q,\omega}(q_z) f(q_z) + \kappa^z_{q,\omega}(q_z) f(0) \right]
~,
\end{eqnarray}
which is rewritten as
\begin{eqnarray} \label{eq:WH_3}
g(q_z) - [ q_{{\rm TF},M}^2 \kappa^z_{q,\omega}(q_z) + i q_z ]  f(0) + f'(0)  = - \big[q_z^2 + q^2 - q_{{\rm TF},M}^2 \kappa_{q,\omega}(q_z) \big] f(q_z)
~,
\end{eqnarray}

We now want to apply the theorem of Eq.~(\ref{eq:splitting_product}) to the right-hand side of Eq.~(\ref{eq:WH_3}). In order to do so, we have to consider the roots of the equation
\begin{eqnarray} \label{eq:WH_4}
q_z^2 + q^2 - q_{{\rm TF},M}^2 \kappa_{q,\omega}(q_z) = 0
~,
\end{eqnarray}
Since $\kappa_{q,\omega}(q_z)$ is even in $q_z$, the solutions of Eq.~(\ref{eq:WH_4}) in the strips $|\Im q_z| < {\bar q}_z$ are denoted by $\pm q_{z,i}$ ($i = 1,\ldots,r$). Without loss of generality, we assume that $\Im(q_{z,i}) \geq 0$ for all $i = 1,\ldots,r$. We assume $\Im(q_{z,i})\neq 0$ for all $i = 1,\ldots,r$, and we order the roots such that $\Im(q_{z,1})< \ldots<\Im(q_{z,r})$. 
We define
\begin{eqnarray}
&& 
P(q_z) = \left\{
\begin{array}{ll}
(q_z^2 - q_{z,1}^2) \cdots (q_z^2 - q_{z,r}^2) & {\rm if}~ r\neq 0
\\
1 & {\rm if}~ r=0
\end{array}
\right.
~,
\nonumber\\
&&
\tau(q_z) = \frac{(q_z^2 + {\bar q}_z^2)^{r-1}}{P(q_z)} \big[ (q_z^2 + q^2) - q_{{\rm TF},M}^2 \kappa_{q,\omega}(q_z) \big]
~.
\end{eqnarray}
Note that with this definition $\tau(q_z)$ is analytic and different from zero in the entire strip $-{\bar q}_z<\Im(q_z)<{\bar q}_{z}$, and goes as $\tau(q_z) \to 1$ for $|q_z| \to \infty$. Therefore we can apply the result of Eq.~(\ref{eq:splitting_product}) and decompose it as $\tau(q_z) = \tau_+(q_z)/\tau_-(q_z)$, where [Eq.~(\ref{eq:def_B_plus_minus})]
\begin{eqnarray} \label{eq:tau_pm_def}
\tau_\pm(q_z) = \exp \left[ \int_{i c_\pm - \infty}^{i c_\pm + \infty} \frac{d{\tilde q}_z}{2\pi i} \frac{\ln \tau({\tilde q}_z)}{{\tilde q}_z-q_z} \right]
~,
\end{eqnarray}
where $|c_\pm|<{\bar q}_z$. $\tau_+(q_z)$ [$\tau_-(q_z)$] is independent of the choice of $c_+$ ($c_-$). Now, sending $c_+\to -{\bar q}_z$ ($c_-\to {\bar q}_z$), we obtain that the so defined $\tau_+(q_z)$ [$\tau_-(q_z)$] is regular and bounded for $\Im q_z > -{\bar q}_z$ ($\Im q_z < {\bar q}_z$). We now define
\begin{eqnarray} \label{eq:Phi_pm_def}
\Phi_\pm(q_z) = (q_z \pm i {\bar q}_z)^{\mp(r-1)} \tau_{\pm}(q_z)
~,
\end{eqnarray}
and we note that $\Phi_+(q_z)$ [$\Phi_-(q_z)$] is still a regular function for $\Im q_z > -{\bar q}_z$ ($\Im q_z < {\bar q}_z$). With this definition Eq.~(\ref{eq:WH_3}) becomes
\begin{eqnarray} \label{eq:WH_5}
\frac{P_-(q_z) f(q_z)}{\Phi_-(q_z)}
+ \frac{ g(q_z) - i q_z  f(0) + f'(0) }{P_+(q_z)\Phi_+(q_z)}
- \frac{q_{{\rm TF},M}^2 \kappa^z_{q,\omega}(q_z) f(0)}{P_+(q_z) \Phi_+(q_z)} = 0
~.
\end{eqnarray}
Here we introduced
\begin{eqnarray}
P_\pm(q_z) = (q_z \pm q_{z,1})\cdots(q_z \pm q_{z,r})
~,
\end{eqnarray}
whose inverse is analytic in the upper (lower) half of the complex plane.
We now note that the first term of Eq.~(\ref{eq:WH_5}) is regular for $\Im (q_z) < {\bar q}_z$, while the second one is regular for $\Im (q_z) > -{\bar q}_z$. The third term, however, is regular only in the strip $-q_{z,1}<\Im(q_z) <{\bar q}_z$. The aim now is to split into two terms $\Psi_+(q_z)$ and $\Psi_-(q_z)$, using Eq.~(\ref{eq:splitting_sum}). Therefore, we define
\begin{eqnarray}
\Psi_+(q_z) + \Psi_-(q_z) = \frac{\kappa^z_{q,\omega}(q_z)}{P_+(q_z) \Phi_+(q_z)}
~,
\end{eqnarray}
where $\Psi_-(q_z)$ is regular for $\Im (q_z) < {\bar q}_z$ and $\Psi_+(q_z)$ is regular for $\Im (q_z) > -q_{z,1}$ [or $\Im (q_z) > -{\bar q}_{z}$, if Eq.~(\ref{eq:WH_4}) has no zeros]. The two functions are defined as
\begin{eqnarray}
\Psi_\pm(q_z) = \pm \int_{i c_\pm -\infty}^{i c_\pm + \infty} \frac{d{\tilde q}_z}{2\pi i} \frac{\kappa^z_{q,\omega}({\tilde q}_z)}{({\tilde q}_z-q_z)P_+(q_z) \Phi_+(q_z)}
~.
\end{eqnarray}
Where $-q_{z,1}< c_+ < \Im(q_z) <  c_- < {\bar q}_z$. Note that the integrand goes to zero faster than $1/|{\tilde q}_z|$ for $|{\tilde q}_z| \to \infty$, and the hypothesis of the theorem are therefore satisfied.
With this definition Eq.~(\ref{eq:WH_5}) is re-arranged as
\begin{eqnarray} \label{eq:WH_6}
\frac{P_-(q_z) f(q_z)}{\Phi_-(q_z)}
- q_{{\rm TF},M}^2 f(0) \Psi_-(q_z)
=
q_{{\rm TF},M}^2 f(0) \Psi_+(q_z) 
-\frac{ g(q_z) - i q_z  f(0) + f'(0) }{P_+(q_z)\Phi_+(q_z)} 
~.
\end{eqnarray}
Now the left-hand side is regular for $\Im (q_z) < {\bar q}_z$, while the right-hand side is regular for $\Im (q_z) > -q_{z,1}$, and they coincide on the strip $-q_{z,1}< \Im (q_z) < {\bar q}_z$. Therefore, together they define a function $\psi(q_z)$ analytic in all the complex plane. Note that $g(q_z) \sim 1/|q_z|$ and $\Psi_\pm(q_z) \sim 1/|q_z|$ for $|q_z| \to \infty$, while $\tau_{\pm}(q_z)$ are bounded. The left-hand side goes as $q_z^0$ as $|q_z| \to \infty$, and it is therefore equal to a constant $Q$ to be determined. Therefore
\begin{eqnarray} \label{eq:WH_7}
f(q_z) = \frac{\Phi_-(q_z)[Q + q_{{\rm TF},M}^2 f(0) \Psi_-(q_z)]}{P_-(q_z)}
~.
\end{eqnarray}
The overall constant is determined from the large-$q_z$ behavior of $f(q_z)$ given in Eq.~(\ref{eq:asymptotics_f}), which implies that $Q = i f(0)$, and therefore
\begin{eqnarray} \label{eq:WH_8}
f(q_z) = if(0) \frac{\Phi_-(q_z)[1 - i q_{{\rm TF},M}^2 \Psi_-(q_z)]}{P_-(q_z)}
~.
\end{eqnarray}

We now calculate $Z_{q,\omega} \equiv qf(0)/f'(0)$. The value of $f'(0)$ is obtained by taking the following limit:
\begin{eqnarray} \label{eq:f_prime_def}
f'(0) &=& \lim_{q_z\to \infty} \big[ i q_z f(0) -q_z^2 f(q_z)  \big]
\nonumber\\
&=& 
i  f(0) \lim_{q_z\to \infty} \left[ q_z - q_z \frac{\tau_-(q_z) (1 - i {\bar q}_z/q_z)^{1-r} [1 - q_{{\rm TF},M}^2 \Psi_-(q_z)]}{\displaystyle\prod_{i=1}^r (1 + q_{z,i}/q_z)} \right]
\nonumber\\
&=&
i f(0)\big[i {\nu'} + i \nu +i (1-r) {\bar q}_z + q_{z,1} + \ldots + q_{z,r} \big]
~,
\end{eqnarray}
where we set $\tau_-(q_z) = 1 - i \nu/q_z$ and we used that $\Psi_-(q_z) = {\nu}'/(q_{{\rm TF},M}^2 q_z)$ for $|q_z| \to \infty$.
$\nu$ is determined from 
\begin{eqnarray} \label{eq:nu_def}
\nu &=& -\int_{-\infty}^{\infty}\frac{dq_z}{2\pi} \ln\tau(q_z)
\nonumber\\
&=&
- \int_{0}^{\infty}\frac{dq_z}{2\pi} \ln\big[\tau(q_z)\tau(-q_z)\big]
\nonumber\\
&=&
-\int_{0}^{\infty}\frac{dq_z}{\pi} \ln\left[ \frac{(q_z^2 + {\bar q}_z^2)^{r}}{P(q_z)} \frac{q_z^2 + q^2 - q_{{\rm TF},M}^2 \kappa_{q,\omega}(q_z)}{q_z^2 + {\bar q}_z^2} \right] 
~.
\end{eqnarray}
where we took $c\to 0^+$ in Eq.~(\ref{eq:tau_pm_def}). Therefore, the integral on the last line of Eq.~(\ref{eq:nu_def}) is intended to be evaluated for $z$ infinitesimally above the real axis. 
In a similar way (we send $c_\pm\to 0^\mp$ in the integration boundaries)
\begin{eqnarray}
\nu' =  q_{{\rm TF},M}^2 \int_{-\infty}^{+\infty} \frac{d{\tilde q}_z}{2\pi i} \frac{\kappa^z_{q,\omega}({\tilde q}_z)}{P_+(q_z) \Phi_+(q_z)}
~,
\end{eqnarray}
%


Finally, we get
\begin{eqnarray}
Z_{q,\omega} = - \frac{q}{{\nu'} + \nu + (1-r) {\bar q}_z -i(q_{z,1} + \ldots + q_{z,r})}
~.
\end{eqnarray}

\end{document}